\documentclass[pra,aps,twocolumn]{revtex4}
\usepackage{amssymb}
\usepackage{latexsym}
\usepackage{endnotes}
\newcommand{\Z}{{\mathbb Z}}

\newcommand{\R}{{\mathbb R}}

\def\be{\begin{equation}}
\def\ee{\end{equation}}
\def\bea{\begin{eqnarray}}
\def\eea{\end{eqnarray}}

\def\d{{\,\rm d}}

\def\vfi{\varphi}
\def\fihat{\hat{\varphi}}
\def\0{{\bf 0}}
\def\a{{\bf a}}
\def\b{{\bf b}}
\def\k{{\bf k}}

\def\n{{\bf n}}
\def\r{{\bf r}}
\def\p{{\bf p}}

\def\x{{\bf x}}
\def\y{{\bf y}}

\def\h2m{\frac{\hbar^2}{2m}}
\def\p0{{P_{\beta H^0_N}}}

\begin{document}

\title{
\large\bf Crystalline ground states for classical particles}
\author{Andr\'as S\"ut\H o \\
Research Institute for Solid State Physics and Optics,
P. O. B. 49, H-1525 Budapest, Hungary}
\thispagestyle{empty}
\begin{abstract}
\noindent
Pair interactions whose Fourier transform is nonnegative and vanishes above a
wave number $K_0$ are shown to give rise to periodic and aperiodic infinite volume ground state
configurations (GSCs) in any dimension $d$. A typical three dimensional example
is an interaction of asymptotic form $\cos K_0 r/r^4$.
The result is obtained for densities
$\rho\geq \rho_d$ where $\rho_1=K_0/2\pi$, $\rho_2=(\sqrt{3}/8)(K_0/\pi)^2$
and $\rho_3=(1/8\sqrt{2})(K_0/\pi)^3$. At $\rho_d$ there is a unique
periodic GSC which is the uniform chain, the triangular lattice and
the bcc lattice for $d=1,2,3$, respectively. For $\rho>\rho_d$ the GSC is non-unique
and the degeneracy is continuous: {\em Any} periodic configuration of density $\rho$ with all
reciprocal lattice vectors not smaller than $K_0$, and any union of such configurations,
is a GSC. The fcc lattice is a GSC only for $\rho\geq(1/6\sqrt{3})(K_0/\pi)^3$.

\vspace{2mm}
\end{abstract}
\maketitle

Crystallization of fluids is the paradigm of a continuous
symmetry breaking. Its conceptual importance has long been recognized \cite{Pei,And},
yet its derivation from first principles is still missing.
Nature and computers can easily
produce it, but a theoretical understanding of the emergence of a periodic
order in continuous space as a result of a translation invariant interaction
appears to be particularly hard. The very first step along the way to a proof of a
phase transition is to show that such interactions do have periodic ground states.
This much more modest program has been
advancing also very slowly, and for a long time the results were limited
mainly to one
dimension \cite{Kunz,VNR,Rad1}. A ground state configuration (GSC) is a minimizer,
in a sense to be defined precisely, of the interaction energy.
It is only recently that the mere existence, without characterization, of
infinite volume GSC was proved for a class of interactions in all dimensions
\cite{Rad}. The first two-dimensional example
of an interaction giving
rise to the triangular lattice as a GSC is even more recent \cite{The}.
The main concern of
this paper is to provide examples of ground state ordering in three dimensions. The system we study
is composed of identical classical particles interacting via
pair interactions whose Fourier transform is nonnegative and decays to zero at
a $K_0<\infty$.
Our results, although not predictive below a dimension-dependent density
$\rho_d\sim K_0^d$, are rather unexpected.
At $\rho_d$ a Bravais lattice (bcc for $d=3$) is the unique
periodic GSC.
At higher densities the set of GSC is continuously degenerate:
within certain limits, volume-preserving deformations
can be done on every GSC without cost of energy, thus yielding other GSC.
The degeneracy increases with the density in the sense that
compressing any GSC results in a GSC of a higher density that can further be deformed.
We can understand this proliferation of GSC
as a consequence of the insensitivity of the interaction to details
on a length scale shorter than $K_0^{-1}$. That
bcc lattice can be more stable than fcc, should not surprise the reader.
At equal densities the fcc nearest neighbor
distance is slightly larger than the bcc one. For a purely repulsive interaction,
the fcc lattice is expected to be
more stable; for a partly attractive interaction at some densities the bcc lattice can
have a lower energy.

{\em Definitions and notations.|}
We shall deal with translation invariant symmetric pair interactions
$\vfi(\r-\r')=\vfi(\r'-\r)$. Rotation invariance will not be supposed. The
$N$-particle configurations ($N\leq\infty$)
are subsets of $N$ points of $\R^d$ and
will be denoted by Latin capitals $B$, $R$, $X$, $Y$.
Infinite configurations with a bounded local density will only be considered.
The number of points in
$R$ will be denoted by $N_R$. If $R$ is a finite configuration, the interaction
energy of $R$ is
$
U(R)=\frac{1}{2}\sum_{\r,\r'\in R, \r\neq\r'}\vfi(\r-\r').
$
We will assume that $\vfi$ and
$
\fihat(\k)=\int_{\R^d}\vfi(\r)e^{-i\k\cdot\r}\d\r
$
are absolutely integrable on $\R^d$.
This ensures that both $\fihat$ and
$
\vfi(\r)=(2\pi)^{-d}\int_{\R^d}\fihat(\k)e^{i\k\cdot\r}\d\k
$
are continuous functions decaying at the infinity \cite{Rud}.
In the theorem below $\fihat\geq 0$ implies $U(R)\geq -\vfi(0)N_R/2$,
hence, stability \cite{Ru}.
Let $R$ and $X$ be a finite and an infinite configuration, respectively.
The energy of $R$ subject to the field created by $X$ is given by
\be\label{rel}
U(R|X)=U(R)+\sum_{\r\in R}\sum_{\x\in X}\vfi(\r-\x).
\ee
Fix a real $\mu$.
An infinite configuration $X$ is called a {\em grand canonical
ground state configuration for
chemical potential $\mu$}
($\mu$GSC) if it is stable against bounded perturbations, i.e., if
for any bounded domain $\Lambda$ and any $R$
\be\label{GSC}
U(R\cap\Lambda|X\setminus\Lambda)-\mu N_{R\cap\Lambda}
\geq U(X\cap\Lambda|X\setminus\Lambda)-\mu N_{X\cap\Lambda}
\ee
where $X\setminus\Lambda$ is the set of points of $X$ outside $\Lambda$.
Because every bounded domain
is in a parallelepiped, these suffice to be considered.
$X$ is a {\em canonical ground state configuration} (GSC) if
(\ref{GSC}) holds for every $R$ such that $N_{R\cap\Lambda}=N_{X\cap\Lambda}$.
Thus, any $\mu$GSC is a GSC. If $\vfi$ is superstable \cite{Ru}, the relation
$\mu\mapsto\rho$ (density) is invertible and any GSC is expected to be
a $\mu$GSC for a suitable $\mu$.
Local stability in the sense of Eq.~(\ref{GSC}) implies global stability, i.e., a
GSC minimizes the energy density at the given density,
cf. \cite{Sew} and the end of this paper.

A Bravais lattice $B=\{\sum_{\alpha=1}^dn_\alpha \a_\alpha | \n\in\Z^d\}$
will be regarded as an infinite
configuration. Here $\a_\alpha$ are linearly independent vectors and
$\n=(n_1,\ldots,n_d)$ is a $d$-dimensional integer. Any periodic configuration $X$
can be written as
$
X=\cup_{j=1}^J(B+\y_j)
$
where $B$ is some Bravais lattice and $B+\y$ is $B$ shifted by the vector $\y$.
For a given $X$,
$B$ is non-unique and we shall choose it so that $J$ be minimum.
Then, we call $X$ a $B$-periodic
configuration. The reciprocal lattice is
$B^*=\{\sum n_\alpha\b_\alpha | \n\in\Z^d\}$ where
$\a_\alpha\cdot\b_\beta=2\pi\delta_{\alpha\beta}$. The nearest neighbor distance
(the length of the shortest nonzero vector) in $B^*$ will be denoted by $q_{B^*}$.
This is related to the density via $\rho(B)=c_{\rm type} (q_{B^*})^d$,
where $c_{\rm type}$ is determined by the aspect ratios and angles of the primitive cell of $B$.
Let $\Lambda$ be the parallelepiped spanned by the vectors $L_\alpha\a_\alpha$,
$\Lambda=\{\sum x_\alpha \a_\alpha | 0\leq x_\alpha<L_\alpha\}$. We shall
take $L_\alpha$ to be positive integers; then $\Lambda$
is a period cell for $B$-periodic configurations, and
the dual lattice
$\Lambda^*=\{\sum(n_\alpha/L_\alpha)\b_\alpha | \n\in\Z^d\}$ contains $B^*$.
Next, we define the {\em periodized} pair interaction
\be\label{phiper}
\vfi_\Lambda(\r)=\sum_{\n\in\Z^d}\vfi\left(\r+\sum n_\alpha L_\alpha\a_\alpha\right)
\ee
and, for $R$ in $\Lambda$, the periodized interaction energy
$U_\Lambda(R)=\frac{1}{2}\sum_{\r,\r'\in R,\r\neq\r'}\vfi_\Lambda(\r-\r')$.
The sum defining $\vfi_\Lambda(\r)$ is
uniformly convergent, therefore $\vfi_\Lambda(\r)$ is continuous and tends to
$\vfi(\r)$ as each $L_\alpha$ tends to infinity.
Finally, let
\be
\mu_\Lambda=\mu+\frac{1}{2}[\vfi(0)-V(\Lambda)^{-1}\sum_{\k\in\Lambda^*}
\fihat(\k)]
\ee
where $V(\Lambda)$ is the volume of $\Lambda$. $\mu_\Lambda$ tends to $\mu$ as
$\Lambda$ increases to $\R^d$.

The main observation leading to the result presented below is as follows. Let $\r_1,\ldots,
\r_N$ be any finite configuration. If $\fihat(\k)\geq 0$ and is zero for $|\k|\geq K_0$ then
\be\label{lite}
\sum_{i,j=1}^N\varphi(\r_i-\r_j)=(2\pi)^{-d}\int_{|\k|<K_0}\fihat(\k)|\sum_{j=1}^N
e^{i\k\cdot\r_j}|^2\d\k\geq 0.
\ee
If we use periodic boundary conditions in a box containing $\r_j$, the integral is to be
replaced by a sum. The $\k=\0$ term of this sum is structure-independent (for $N$
fixed) and the rest is
non-negative. Hence, any structure making the rest vanish is a GSC. But that is exactly what
periodic structures accomplish, provided their shortest reciprocal lattice vector is outside the
$|\k|=K_0$ sphere.

{\em THEOREM. Let both $\vfi$ and $\fihat$ be absolutely integrable in $\R^d$,
$\fihat(\k)=\fihat(-\k)\geq 0$ and $\fihat(\k)=0$ for
$|\k|\geq K_0$. We have the following results.\\
(i) Let $B$ be a Bravais lattice with $q_{B^*}\geq K_0$. Then every
$B$-periodic configuration $X$ is a GSC, and its energy per volume,
$e(X)=\frac{1}{2}\rho[\rho\fihat(0)-\vfi(0)]$, is minimum for the density $\rho=\rho(X)$.
On every period cell $\Lambda$, $X\cap\Lambda$
minimizes $U_\Lambda(R)$ for fixed $N_R=N_{X\cap\Lambda}$.
Also, $X$ creates a force-free field on test particles, i.e.,
$U(\r|X)$ is independent of $\r$.
Any union of GSCs of the above type is a GSC (that can be aperiodic).
\\
(ii) There is a smallest density $\rho_d$ at which
$q_{B^*}=K_0$ holds for a single Bravais lattice. If $\fihat(\k)>0$ for $0<|\k|<K_0$,
this $B$ is the only periodic GSC.
$\rho_1=K_0/2\pi$, $\rho_2=(\sqrt{3}/8)(K_0/\pi)^2$
and $\rho_3=(1/8\sqrt{2})(K_0/\pi)^3$, and the respective GSC are
the uniform chain, the triangular lattice and
the bcc lattice. 
The fcc, simple hexagonal (sh), simple cubic (sc)
and hcp lattices are GSCs at and above the respective densities
$\rho_{\rm fcc}=(1/6\sqrt{3})(K_0/\pi)^3$,
$\rho_{\rm sh}=(\sqrt{3}/16)(K_0/\pi)^3$, $\rho_{\rm sc}=(1/8)(K_0/\pi)^3$ and
$\rho_{\rm hcp}=(4/3\sqrt{3})(K_0/\pi)^3$.\\
(iii) Suppose, in addition, that $\fihat(0)>0$ and
$[\mu+\frac{1}{2}\vfi(0)]/\fihat(0)\geq\rho_d$.
Then any $B$-periodic
configuration $X$ such that $q_{B^*}\geq K_0$ (equivalently, $\rho(B)\geq\rho_d$)
and $\rho(X)=[\mu+\frac{1}{2}\vfi(0)]/\fihat(0)$ is a
$\mu$GSC, and its energy density, $e_\mu(X)=e(X)-\mu\rho(X)=
-\frac{1}{2}\rho(X)^2\fihat(0)$, is minimum for the given $\mu$.
If $\Lambda$ is a period cell,
$X\cap\Lambda$ minimizes $U_\Lambda(R)-\mu_\Lambda N_R$.}

Because all the moments of $\fihat$ are finite, $\vfi$ is infinitely differentiable.
A hard-core interaction can be added to $\vfi$ provided that the
close-packing density is larger than $\rho_d$. Its only role is to restrict the
set of allowed configurations.
The set of GSC is reminiscent of a compressible fluid.
The canonical and grand canonical ground state energy densities,
\be\label{e1}
e_\rho=\frac{\rho}{2}[\rho\fihat(0)-\vfi(0)],\quad
e_\mu=-\frac{1}{2\fihat(0)}\left[\mu+\frac{\vfi(0)}{2}\right]^2
\ee
are Legendre transforms of each other,
\be\label{Legendre}
e_\rho=\max_\mu\{e_\mu+\mu\rho\},\quad
e_\mu=\min_\rho\{e_\rho-\mu\rho\}.
\ee
The $\rho$-dependence of $e_\rho$ shows that the interaction is stable if
$\fihat(0)\geq 0$ and is superstable \cite {Ru} if $\fihat(0)>0$.
In the second case (\ref{Legendre})
expresses the equivalence of the canonical and grand canonical ensembles.
The corresponding density and chemical potential satisfy the equation
\be
\mu+\vfi(0)/2-\fihat(0)\rho=0.
\ee
This linear relation breaks down for small densities because $\mu$ has to tend to $-\infty$
as $\rho$ approaches zero.
This implies a nonanalyticity, presumably at $\rho_d$.

{\em Examples.|} Based on a result on Fourier transforms \cite{RS},
we can obtain fast-decaying (but not finite-range)
interactions satisfying the conditions of the theorem. Take any locally integrable
real function $g(\k)=g(-\k)\geq 0$, fix an $\varepsilon>0$ and define
$
\fihat(\k)=\int_{|\k'|<K_0-\varepsilon}g(\k')\eta_\varepsilon(\k-\k')\d\k'
$
where
$
\eta_\varepsilon(\k)=  
\exp\left[-(1-k^2/\varepsilon^2)^{-1}\right]
$  
if $k<\varepsilon$ and 0 otherwise.
This $\fihat$ is infinitely differentiable.
By inverse-Fourier transforming it we find $\vfi$ to
decay faster than algebraically.
More interesting are the long-range interactions. They can be obtained by choosing
$\fihat$ to be only finitely many times differentiable (namely, at $|\k|=K_0$).
For example, in one
dimension $\fihat(k)=K_0-|k|$ for $|k|\leq K_0$ yields
$\vfi(x)=(1-\cos K_0 x)/\pi x^2$. In three dimensions rotation invariant examples
can be obtained by starting with a function $f(k)\geq 0$ such that
$f$ is three times continuously differentiable and $f(K_0)=f'(K_0)=0$.
Then, defining $\fihat(\k)=f(|\k|)$ for
$|\k|\leq K_0$, by partial integration
\be
\vfi(\r)=\frac{1}{2\pi^2 r^4}\left\{\left[(kf)''\cos kr\right]_0^{K_0}
-\int_0^{K_0}(kf)'''\cos kr \d k\right\}.
\ee
For instance,
$f(k)=\pi^2(k+z)(k+\overline{z})(k-K_0)^2$ with $z=(K_0/10)(1+3i)$ gives
$\vfi(\r)=(13/10)K_0^3\cos K_0r/r^4 + O(1/r^5)$.
The higher order terms make $\vfi$ finite at the origin.
One can verify that $\vfi(0)/2\fihat(0)>\rho_3$, so
this interaction has a continuous
family of inequivalent $\mu$GSC at $\mu=0$.
Also, it has the bcc lattice as the unique periodic GSC at density
$(1/8\sqrt{2})(K_0/\pi)^3$.

{\em LEMMA. Let
$\Lambda$ be
a parallelepiped spanned by the vectors $L_\alpha\a_\alpha$. Then
\be\label{Poisson}
\varphi_\Lambda(\r)=V(\Lambda)^{-1}\sum_{\k\in\Lambda^*}\fihat(\k)e^{i\k\cdot\r}.
\ee
}

In one dimension, for $r=0$, Eq.~(\ref{Poisson}) reduces
to the Poisson summation formula \cite{Rud}.

{\em Proof of the lemma.|}
The Fourier coefficients of $\vfi_\Lambda$ are
$\int_\Lambda\vfi_\Lambda(\r)e^{-i\k\cdot\r}\d\r$. Substituting the sum (\ref{phiper}) for
$\vfi_\Lambda(\r)$, integrating by terms, and resumming,
we obtain $\fihat(\k)$. The series on the right is 
a continuous periodic function whose Fourier coefficients
are also $\fihat(\k)$. Because of the completeness of the system
$\{e^{i\k\cdot\r}|\k\in\Lambda^*\}$ in the Banach space of integrable functions
on $\Lambda$, equality for all $\r$ follows.

{\em Proof of the theorem.| (i),(iii)} Consider the periodic configuration
$
X=\cup_{j=1}^J(B+\y_j)
$
and let $R$ be obtained from $X$ by a bounded perturbation. Take a period
parallelepiped $\Lambda$ of $X$ large enough to contain the perturbed part, that is,
$R=X$ outside $\Lambda$.
Recall that $\Lambda^*$ contains $B^*$ as a part. For a $\k$ in $\Lambda^*$,
\be\label{esum}
\sum_{\x\in X\cap\Lambda}e^{-i\k\cdot\x}=\chi_{B^*}(\k)\ N_{B\cap\Lambda}
\sum_{j=1}^Je^{-i\k\cdot\y_j}
\ee
where $\chi_{B^*}(\k)=1$ if $\k$ is in $B^*$ and 0 otherwise. Using the lemma
and Eq.~(\ref{esum}), after some computation one finds
\begin{widetext}
\bea\label{Urel}
U(R\cap\Lambda|X\setminus\Lambda)=-\frac{1}{2}\,\vfi(0)\,N_{R\cap\Lambda}
+\int\frac{\fihat(\k)}{2(2\pi)^{d}}   
\left[\left|\sum_{\r\in R\cap\Lambda}e^{i\k\cdot\r}\,\right|\,^{^2}-
2\sum_{\r\in R\cap\Lambda}e^{i\k\cdot\r}\sum_{\x\in X\cap\Lambda}e^{-i\k\cdot\x}
\right]\d\k    \nonumber\\
+ \rho(B)\sum_{\k\in B^*}\fihat(\k)\sum_{j=1}^Je^{-i\k\cdot\y_j}
\sum_{\r\in R\cap\Lambda}e^{i\k\cdot\r}.
\eea
Here $\rho(B)=N_{B\cap\Lambda}/V(\Lambda)$, the density of $B$. We can obtain
$U(X\cap\Lambda|X\setminus\Lambda)$ from Eq.~(\ref{Urel}) if
we replace $R$ by $X$ and reapply Eq.~(\ref{esum}). Finally, the condition
(\ref{GSC}) reads
\bea\label{main}
\frac{1}{2}(2\pi)^{-d}\int\fihat(\k) 
\left|\sum_{\x\in X\cap\Lambda}e^{i\k\cdot\x}-
\sum_{\r\in R\cap\Lambda}e^{i\k\cdot\r}\,\right|\,^{^2} \ \d\k  
\geq\, (N_{R\cap\Lambda}-N_{X\cap\Lambda})\ [\mu+\vfi(0)/2-\rho(X)\fihat(0)]\nonumber\\
+\,\rho(B)\sum_{\0\neq\k\in B^*}\fihat(\k)   
\left[N_{B\cap\Lambda}\ |\sum_{j=1}^Je^{i\k\cdot\y_j}\,|\,^2-
\sum_{j=1}^Je^{-i\k\cdot\y_j}
\sum_{\r\in R\cap\Lambda}e^{i\k\cdot\r}\right].
\eea
\end{widetext}
Here we used $N_{X\cap\Lambda}=JN_{B\cap\Lambda}$, $\rho(X)=J\rho(B)$.
If $\fihat(\k)\geq 0$ and is zero for $|\k|\geq q_{B^*}$,
the left member is nonnegative, and the sum multiplying $\rho(B)$ vanishes for all $R$.
These and $N_{R\cap\Lambda}=N_{X\cap\Lambda}$ ensure that (\ref{main}) holds true, so that $X$ is
a GSC. If, moreover,
$\rho(X)=[\mu+\vfi(0)/2]/\fihat(0)$, the first term
of the right member also vanishes for all $R$, and $X$ is a $\mu$GSC.

Choosing any period parallelepiped $\Lambda$, the canonical
energy density of a $B$-periodic $X$ of density $\rho$ is
\bea\label{enden}
e(X)&=&\frac{1}{2V(\Lambda)}\sum_{\x\in X\cap\Lambda}
\sum_{\x\neq\x'\in X}\vfi(\x-\x')   
\\
&=&\frac{1}{2V(\Lambda)}\sum_{\x,\x'\in X\cap\Lambda}
\vfi_\Lambda(\x-\x')-\frac{1}{2}\vfi(0)\rho\nonumber\\
&=&\frac{1}{2}\,\rho(B)^2\sum_{\k\in B^*}\fihat(\k)\,
|\sum_{j=1}^Je^{i\k\cdot\y_j}\,|\,^2 - \frac{1}{2}\vfi(0)\rho.\nonumber
\eea
If $\fihat(\k)=0$ for $|\k|\geq q_{B^*}$ then
$e(X)=\frac{1}{2}\,\fihat(0)\rho^2- \frac{1}{2}\vfi(0)\rho$.
If also $\fihat\geq 0$ then $X$ is a GSC and $e(X)$ 
is the absolute minimum among configurations of density $\rho$.

For $R$ in $\Lambda$ and using again the lemma,
\bea\label{Uper}
U_\Lambda(R)=\frac{1}{2V(\Lambda)}\sum_{\0\neq\k\in \Lambda^*}\fihat(\k)\,\,
|\sum_{\r\in R}e^{i\k\cdot\r}\,|\,^2
\nonumber\\
+\frac{N_R}{2V(\Lambda)}\,\left[\,N_R\,\fihat(0)-\sum_{\k\in\Lambda^*}\fihat(\k)\,\right],
\eea
cf. Eq.~(\ref{lite}).
If $R=X\cap\Lambda$, according to (\ref{esum}), the first sum reduces to
$\0\neq\k\in B^*$
and vanishes completely if $\fihat(\k)= 0$ at $|\k|\geq q_{B^*}$.
When $\fihat\geq 0$,
$X\cap\Lambda$ minimizes $U_\Lambda(R)$ for fixed $N_R=N_{X\cap\Lambda}$ and
$U_\Lambda(R)-\mu_\Lambda N_R$ under the stronger conditions of (iii).
One can also show the opposite implication:
if $X$ is periodic and $X\cap\Lambda_n$ minimizes $U_{\Lambda_n}$ for a
sequence $\Lambda_n\to\R^d$ of period cells then $X$ is a GSC.

With $\Lambda$ to be the primitive unit cell,
\bea
\lefteqn{U(\r|X)=\sum_{\x\in X}\vfi(\r-\x)
=\sum_{\x\in X\cap\Lambda}\vfi_\Lambda(\r-\x)}\\
&&=V(\Lambda)^{-1}\sum_{\x\in X\cap\Lambda}\sum_{\k\in B^*}\fihat(\k)
e^{i\k\cdot(\r-\x)} 
=\rho(X)\fihat(0)\nonumber
\eea
independent of $\r$. This implies that the union of GSCs
is also a GSC that can be aperiodic; see \cite{Su} for details.

{\em (ii)} The one-dimensional case is obvious.
In two dimensions $\rho(B)=(2\pi)^{-2}|\b_1\times\b_2|$,
in three dimensions $\rho(B)=(2\pi)^{-3}|(\b_1\times\b_2)\cdot\b_3|$.
We are going to select $B^*$ by choosing $\b_\alpha$ in such a way that $\rho(B)$
be minimum, on condition that
$
q_{B^*}\equiv\min\{|\sum n_\alpha\b_\alpha|\n\neq 0\}
=K_0$. The resulting Bravais lattice at the given density will be denoted by $B_d$.
In two dimensions, let $\b_1$ be one of the shortest vectors of $B^*$, so
that $b_1=q_{B^*}$.
Consider
the shortest vectors of $B^*$ not collinear with
$\b_1$. These form a star: with $\k$, $-\k$ is also in the set. Choose $\b_2$
among them so that $\b_1\cdot\b_2\geq 0$. In the triangle of sides $b_1$, $b_2$
and $|\b_2-\b_1|$ we have $|\b_2-\b_1|\geq b_2\geq b_1$,
so the largest angle is that
of the vectors $\b_1$ and $\b_2$. Therefore, this angle $\alpha_{12}$
is between $\pi/3$ and
$\pi/2$. Under these restrictions the minimum of $|\b_1\times\b_2|$ is obtained
for $b_2=b_1$ and $\alpha_{12}=\pi/3$. These conditions define a triangular lattice
for $B_2^*$, and also a triangular lattice for $B_2$, so that $\rho_2=\rho_{\rm tr}$.
In three dimensions,
the threshold densities of cubic lattices, at which the length of
their shortest reciprocal
lattice vector is $K_0$, can simply be computed. This gives the order
$\rho_{\rm bcc}<\rho_{\rm fcc}<\rho_{\rm sc}$ and the values presented in the
theorem. For the hexagonal lattice the threshold density depends on $c/a$, and
the minimum is obtained for $c/a=\sqrt{3}/2$. Its value, and that of the hcp
lattice at $c/a=\sqrt{8}/3$ and $J=2$
are given in the theorem. It remains to convince
oneself that the other Bravais lattices cannot give a smaller density.
After some reflection one can conclude that the conditional minimum of
$|(\b_1\times\b_2)\cdot\b_3|$ is attained by choosing $b_1=b_2=b_3=K_0$ and
$\pi/3$ for the three angles between the three pairs of vectors. This specifies
$B_3^*$ as an fcc lattice and $B_3$ as a bcc lattice. Thus, $\rho_3=\rho_{\rm bcc}$.
The last step of the proof of the theorem is to show that at the density
$\rho_d$ no other periodic configuration $X$ can be a GSC. Suppose therefore that
$\rho(X)=\rho_d$. If $X$ is B-periodic but $X\neq B$ then $\rho(B)=\rho_d/J<\rho_d$ and
therefore $q_{B^*}<K_0$. If $X=B$ but $B\neq B_d$, then again $q_{B^*}<K_0$.
Because $\fihat(\k)>0$ for $|\k|< K_0$, at least two nonzero vectors
of $B^*$ give a positive contribution to the energy density (\ref{enden}).
Hence, $e(X)>\frac{1}{2}\rho_d[\rho_d\,\fihat(0)-\vfi(0)]=e(B_d)$ while
$\rho(X)=\rho(B_d)$. Then $X$ is not a GSC because of the following.

Let
$X$ and $Y$ be
two configurations of equal densities but let $e(Y)<e(X)$. Then $X$
cannot be a GSC of $\vfi$. In other words, every GSC minimizes the energy
density at the given number density.
%
Thus, no GSC can be metastable, cf. Ref. \cite{Sew}.
We only outline the proof. 
Take a large domain $\Lambda$ of volume $V$ such that
$N_{X\cap\Lambda}=N_{Y\cap\Lambda}$.
Estimate
$$
U(X\cap\Lambda|X\setminus\Lambda)=U(X\cap\Lambda)+
\sum_{\x\in X\cap\Lambda}\sum_{\x'\in X\setminus\Lambda}\vfi(\x-\x').
$$
The first term is $V(\Lambda)e(X)+o(V)$ while the second is
$o(V)$. Similarly,
$U(Y\cap\Lambda|X\setminus\Lambda)=V(\Lambda)e(Y)+o(V)$.
Thus,
$
U(Y\cap\Lambda|X\setminus\Lambda)<U(X\cap\Lambda|X\setminus\Lambda)
$
for $\Lambda$ large enough.
This concludes the proof of the theorem.

In summary, for a class of translation invariant pair interactions we have proved
the existence of periodic and aperiodic
ground states in classical particle systems.
This result is probably the first of its kind in three dimensions.
Our finding, that ground states of innocent looking RKKY-type interactions
such as $\cos K_0r/r^4$
are continuously deformable by volume preserving transformations without cost of
energy is quite unexpected. In another interpretation this means that
existing crystal structures are stable against
perturbations with interactions described in this paper.

I thank G. Oszl\'anyi, C. Pfister, C. Radin, G. Sewell,
J. S\'olyom and F. Theil for helpful remarks.
This work was supported by OTKA Grants T 042914 and T 043494.

\end{document}